\documentclass[aps,pra,twocolumn,superscriptaddress,nobibnotes,letterpaper]{revtex4-1}

\usepackage[pdftex]{graphicx}
\usepackage{graphicx}
\usepackage{amsmath}
\usepackage{verbatim}
\usepackage{color}
\usepackage{bm}
\usepackage{txfonts}

\usepackage[colorlinks=true, allcolors=blue]{hyperref} %hyperrefs

\newcommand{\FF}{{\cal F}}

\newcommand{\RR}{{\cal R}}

\begin{document}

\title{Proposal for optomechanical quantum teleportation}\thanks{This work was published in \href{https://doi.org/10.1103/PhysRevA.102.032402}{Phys.\ Rev.\ A \textbf{102}, 032402} (2020). The source data for the figures is available at \href{https://doi.org/10.5281/zenodo.4013836}{10.5281/zenodo.4013836}.}

\author{Jie Li}
\affiliation{Kavli Institute of Nanoscience, Department of Quantum Nanoscience, Delft University of Technology, 2628CJ Delft, The Netherlands}

\author{Andreas Wallucks}
\affiliation{Kavli Institute of Nanoscience, Department of Quantum Nanoscience, Delft University of Technology, 2628CJ Delft, The Netherlands}

\author{Rodrigo Benevides}
\affiliation{Kavli Institute of Nanoscience, Department of Quantum Nanoscience, Delft University of Technology, 2628CJ Delft, The Netherlands}
\affiliation{Photonics Research Center, Applied Physics Department, Gleb Wataghin Physics Institute, P.O.\ Box 6165, University of Campinas -- UNICAMP, 13083-970 Campinas, SP, Brazil}

\author{Niccolo Fiaschi}
\affiliation{Kavli Institute of Nanoscience, Department of Quantum Nanoscience, Delft University of Technology, 2628CJ Delft, The Netherlands}

\author{Bas Hensen}
\affiliation{Kavli Institute of Nanoscience, Department of Quantum Nanoscience, Delft University of Technology, 2628CJ Delft, The Netherlands}

\author{Thiago P.\ Mayer Alegre}
\affiliation{Photonics Research Center, Applied Physics Department, Gleb Wataghin Physics Institute, P.O.\ Box 6165, University of Campinas -- UNICAMP, 13083-970 Campinas, SP, Brazil}

\author{Simon Gr\"oblacher}
\email{s.groeblacher@tudelft.nl}
\affiliation{Kavli Institute of Nanoscience, Department of Quantum Nanoscience, Delft University of Technology, 2628CJ Delft, The Netherlands}

\begin{abstract}
We present a novel discrete-variable quantum teleportation scheme using pulsed optomechanics. In our proposal, we demonstrate how an unknown optical input state can be transferred onto the joint state of a pair of mechanical oscillators, without physically interacting with one another. We further analyze how experimental imperfections will affect the fidelity of the teleportation and highlight how our scheme can be realized in current state-of-the-art optomechanical systems.
\end{abstract}

%\date{\today}
\maketitle

\section{Introduction}

The process of quantum teleportation describes the transfer of an unknown input state onto a remote quantum system. First outlined by Bennett et al.~\cite{Bennett1993}, it has since evolved into an active area of research and is now recognized as an important tool for many quantum protocols such as quantum repeaters~\cite{Sangouard2011}, measurement-based quantum computing~\cite{Raussendorf2001}, and fault-tolerant quantum computation~\cite{Gottesman1999}. Experiments have been realized first with photons~\cite{Bouwmeester1997}, later with various systems such as trapped ions~\cite{Riebe2004,Barrett2004}, atomic ensembles~\cite{Sherson2006}, as well as with high-frequency phonons~\cite{Hou2016} and several others~\cite{Pirandola2015}. Over the past few years, optomechanical devices have emerged as an interesting tool to explore quantum phenomena, both from a fundamental perspective, showing the limits of quantum mechanical rules on massive objects~\cite{Aspelmeyer2014}, as well as from an applied view, promising to act as efficient transducers connecting radio-frequency regime qubits to low-loss optical channels~\cite{Forsch2020,Mirhosseini2020}. While continuous variable teleportation using an optomechanical system has been proposed~\cite{Mancini2003,Felicetti2017}, an experimental realization of such a scheme remains beyond reach.\\

Here we present a protocol that will enable the realization of quantum teleportation of an unknown optical input state onto a stationary mechanical quantum memory based on discrete variables in the pulsed regime. The scheme is based on a dual-rail encoding in which the polarization state of a photonic input qubit is teleported onto two mechanical modes. Current state-of-the-art optomechanics devices~\cite{Marinkovic2018} should be able to realize the proposed protocol. The optomechanical interaction is used as a source of Einstein-Podolsky-Rosen (EPR)-type entanglement between the memory and an optical field, after which a successful Bell-state measurement with the input field completes the teleportation. The memory state can be read out on-demand back onto the optical field directly at telecom wavelengths~\cite{Wallucks2020}. With the recent progress in quantum storage in ultra long-lived mechanical oscillators and their on-chip integration, the system represents progress towards scalable quantum networks using quantum repeater schemes. We discuss the expected teleportation fidelities when including experimentally relevant imperfections, such as higher-order excitations from the optomechanical interaction, a finite thermal background on the mechanical mode, optical losses, and nonunity efficiency in the Bell-state detection. We further discuss the feasibility of demonstrating quantum teleportation using weak coherent input states instead of a single-photon source.

\section{Optomechanical interaction}
\label{tools}

We start the discussion with a description of the optomechanical interactions which allow the generation of EPR-type entanglement and retrieve the stored state from the stationary memory after a successful teleportation trial. Our approach is valid for general optomechanical devices, however, the teleportation protocol requires initialization of the modes in the mechanical ground state which to date is most reliably done by cryogenic cooling of nanobeam optomechanical resonators~\cite{Meenehan2015,Riedinger2016,Riedinger2018}. The Hamiltonian for such systems reads~\cite{Aspelmeyer2014}
\begin{equation}
H/\hbar = \omega_c \hat{a}^{\dag} \hat{a} + \omega_m \hat{b}^{\dag} \hat{b} - g_0 \hat{a}^{\dag} \hat{a} (\hat{b}+ \hat{b}^{\dag}) + i (E e^{-i \omega_0 t} \hat{a}^{\dag} - h.c. )  , 
\end{equation}
where $\hat{a}$ and $\omega_c$ ($\hat{b}$ and $\omega_m$) are the annihilation operator and resonant frequency of the cavity (mechanical) mode, respectively, $g_0$ is the single-photon optomechanical coupling rate, and $E=\sqrt{2\kappa P/(\hbar \omega_0)}$ is the coupling between the driving field with frequency $\omega_0$ and power $P$, and the cavity with decay rate $\kappa$ (HWHM). Pulsed optical driving at a laser frequency $\omega_0 = \omega_c \pm \omega_m$ allows for the realization of two different types of optomechanical interactions in the resolved sideband limit $\kappa \ll \omega_m$, namely, the parametric down-conversion and the beamsplitter interaction~\cite{Aspelmeyer2014}. For a blue-detuned pulse, $\omega_b \simeq \omega_c+\omega_m$, the interaction Hamiltonian $H_b \propto \hbar g (\hat{a}^{\dag} \hat{b}^{\dag} +\hat{a} \hat{b})$ results in optomechanical parametric down-conversion, or two-mode squeezing, correlating excitations in the mechanical mode and the Stokes field at the cavity frequency $\omega_c$. Here, $g=g_0\cdot\sqrt{n_c}$ is the effective optomechanical coupling rate, with the intra-cavity photon number $n_c$. This entangling operation is used in the teleportation scheme to generate optomechanical EPR-type states. The success probability is given by $g\tau$, where $\tau$ is the interaction time or, experimentally, the pulse length. Using a coherent laser drive to enable the interaction, double or even multiple scattering events can occur with a probability of the order of $(g\tau)^n$, which we include in the later discussion. For a red-detuned pulse, $\omega_r \simeq \omega_c-\omega_m$, the interaction Hamiltonian $H_r \propto \hbar g (\hat{a}^{\dag} \hat{b} +\hat{a} \hat{b}^{\dag})$ results in an optomechanical state-swap where an anti-Stokes process annihilates a phonon while emitting a photon at cavity resonance. In our scheme, this interaction is used to read out the final mechanical state after a successful teleportation of the initial optical state.

\begin{figure}[t]\label{fig1}
\includegraphics[width=\linewidth]{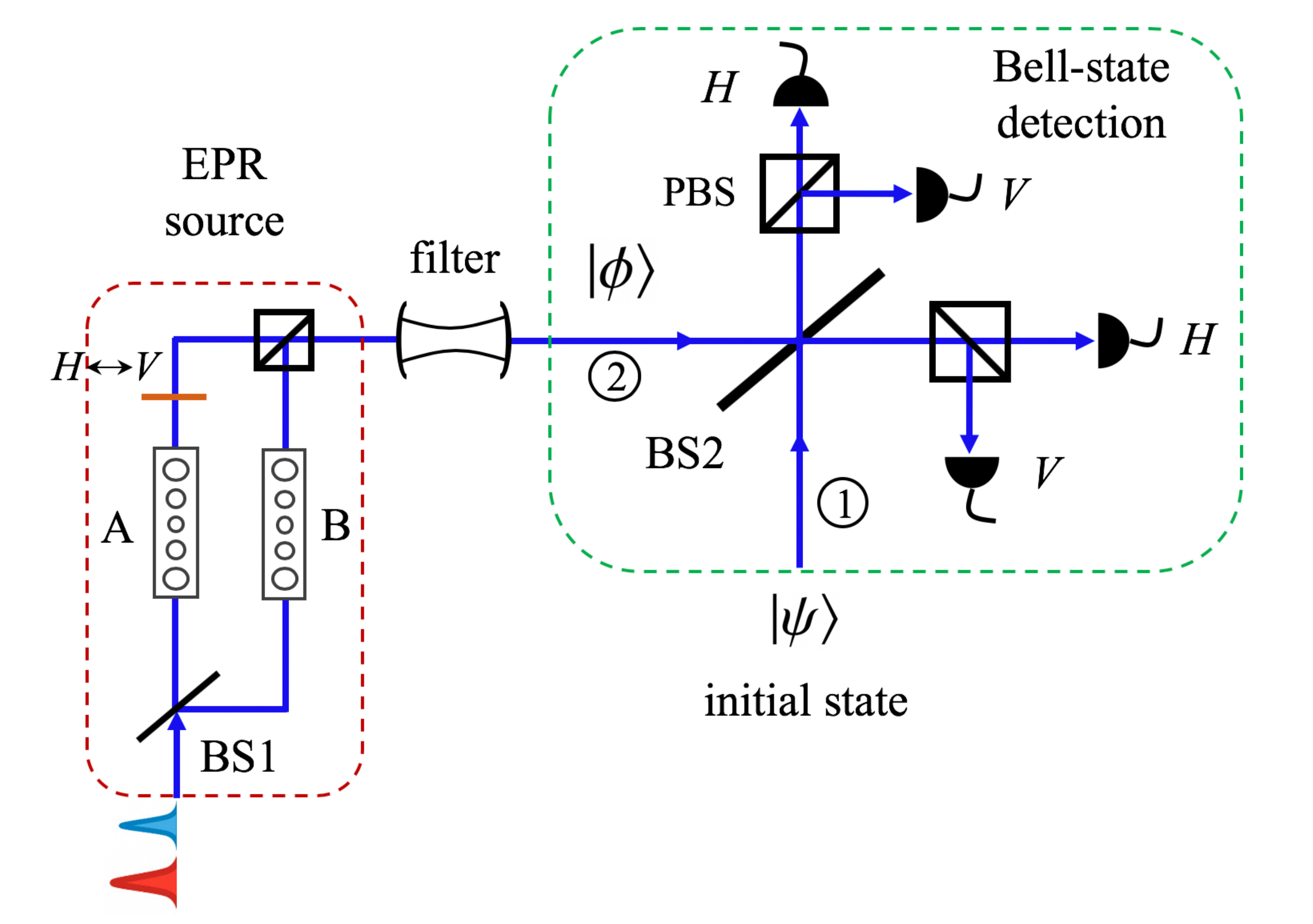}
\caption{Sketch of a proposed setup to implement the teleportation protocol. Two optomechanical devices forming a quantum memory are placed in an optical interferometer in which the outgoing fields are combined in cross polarization. Simultaneous driving of the devices under blue detuning entangles the state of the mechanical modes with optical fields at cavity resonance $\omega_c$. A Bell-state measurement between this Stokes field and an optical state to be teleported $|\psi \rangle$ is performed by mixing the fields on a 50/50 beamsplitter (BS2) and subsequent polarization analysis using two polarizing beamsplitters (PBS). The teleported state is then retrieved from the optomechanical system using red-detuned driving and can be analyzed in the same Bell-state detection setup.}
\label{Fig1}
\end{figure}

\section{Ideal case}
\label{ide}

We now describe the teleportation of a photonic qubit state $|\psi \rangle$ in polarization encoding onto a memory element consisting of two optomechanical devices, as shown in Fig.~\ref{Fig1}. The two resonators are subject to a simultaneous excitation using a blue-detuned drive after which a Bell-state measurement of the input photon with the Stokes field from the optomechanical devices enables the teleportation. For simplicity, we begin our protocol assuming at most single excitations in all relevant modes and subsequently discuss the effects of various expected experimental imperfections. In this ideal case, a blue-detuned single-photon pulse is sent onto a 50/50 beamsplitter (BS1) such that a path-entangled state is generated in its two outputs (which we call path $A$ and $B$), i.e., $\frac{1}{\sqrt{2}} \Big( |01 \rangle + |10 \rangle  \Big)_{AB}$. In each path, the optical pulse interacts with an optomechanical device. By selecting trials with successful scattering events, the two-mode squeezing interaction generates an entangled state of the form
\begin{equation}\label{eq3}
|\phi \rangle=\frac{1}{\sqrt{2}} \Big( |0101 \rangle_{AB HV} + |1010 \rangle_{AB HV} \Big),
\end{equation}
where the subscript $A$ ($B$) denotes the mechanical mode in path $A$ ($B$). Here we distinguish the two optical fields in paths $A$ and $B$ by encoding them in orthogonal polarizations, where $H$ ($V$) denotes horizontal (vertical) polarization. This can be realized by placing a half-wave plate in one of the two optical paths. The optical fields from both paths are then filtered, such that only the photons on resonance with the optomechanical cavities are transmitted to the Bell-state detection setup (cf.\ Fig.~\ref{Fig1}). Note that in order to get the state~\eqref{eq3}, we have assumed the ideal case that the two devices are identical and that mechanical modes are initially in the quantum ground state $| 00 \rangle_{AB}$ and neglected any heating effect due to light absorption.

The initial optical state for the teleportation $|\psi \rangle$ is an arbitrary superposition of two polarization modes of the form
\begin{equation}\label{eq4}
|\psi \rangle= \cos \theta \, |01 \rangle_{HV} + e^{i\varphi}\sin \theta \, |10 \rangle_{HV},
\end{equation}
where $\varphi$ is an arbitrary phase factor, which we choose as $\varphi=0$. Such a state can be prepared by sending a single photon with frequency $\omega_c$ into a polarizing beamsplitter (PBS) with transmission (reflection) coefficient $\cos \theta$ ($\sin \theta$), for example.

\begin{figure}[b]\label{fig2}
\includegraphics[width=\linewidth]{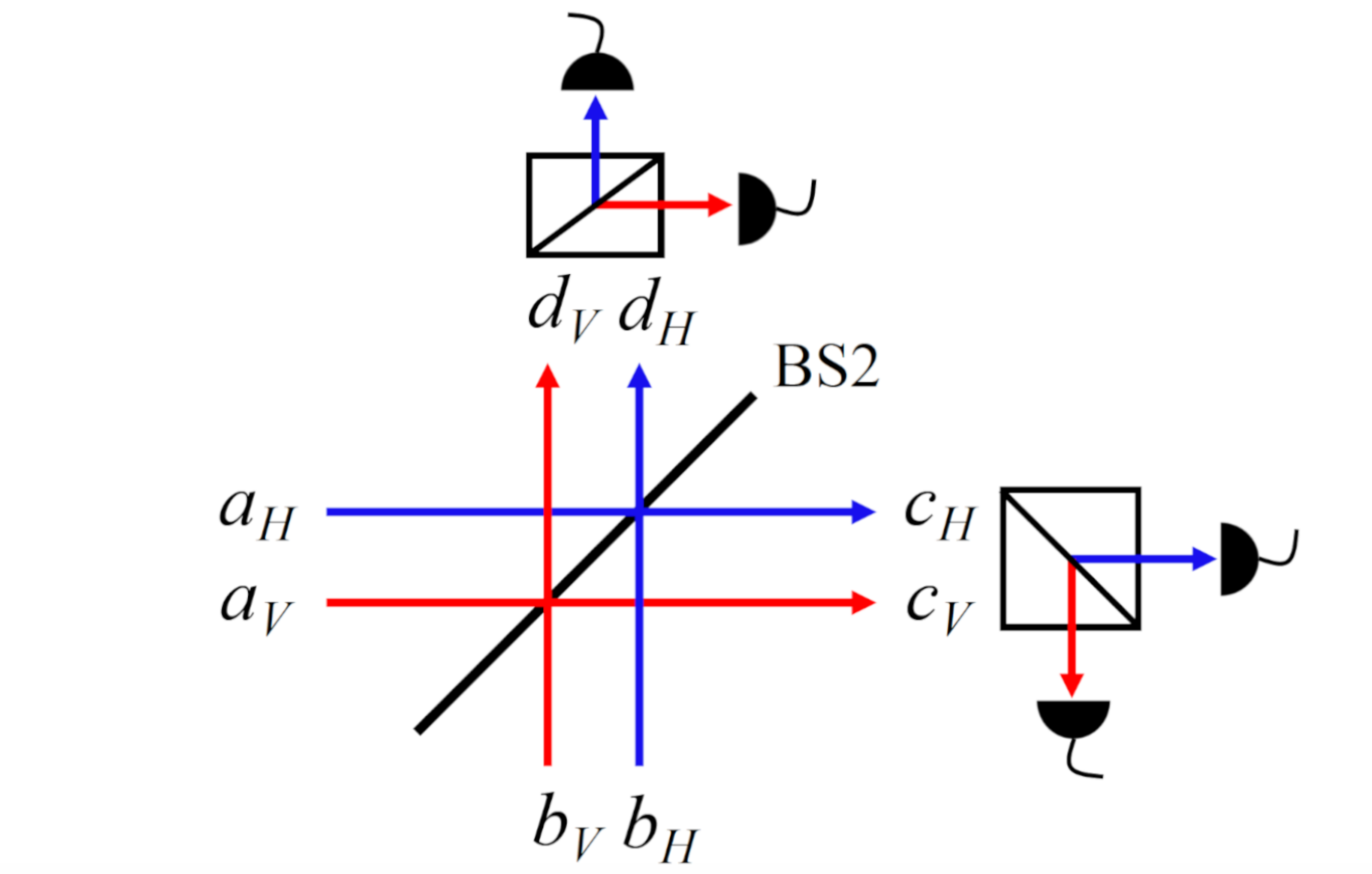}
\caption{Diagram of the Bell-state measurement $\hat{M}_{\pm}$ for two polarization modes, which consists of a 50/50 beamsplitter (BS2) and four single-photon detectors placed after two polarizing beamsplitters (PBS).}
\label{Fig2}
\end{figure}

This input photon is sent to port 1 of the Bell-state detection setup, whereas the output of the interferometer containing the optomechanical devices is sent to port 2. Combining the states of Eqs.~\eqref{eq3} and \eqref{eq4} results in
\begin{equation}\label{eq5}
\begin{split}
|\Phi \rangle =& |\phi \rangle \otimes |\psi \rangle   \\
= & \frac{1}{\sqrt{2}} \Big( \cos\theta \, | 0101 01 \rangle +  \sin \theta \, | 010110  \rangle   \\
 &+  \cos\theta \, | 101001 \rangle  + \sin\theta \, | 101010 \rangle    \Big)_{AB H_2V_2 H_1V_1},
\end{split}
\end{equation}
where the subscripts 1 and 2 are used to distinguish the different input ports (see Fig.~\ref{Fig1}). The Bell-state detection setup consists of a 50/50 beamsplitter BS2 on which the input fields are mixed. Each output is then routed to a polarizing beamsplitter with two single-photon detectors at their respective outputs. A successful teleportation trial is heralded by a coincidence in two of the detectors corresponding to different polarization states, realizing a detection of the input state of Eq.~\eqref{eq5} using
\begin{equation}\label{eq6}
\hat{M}_{\pm}=  \Big( | 0110 \rangle  \pm | 1001 \rangle  \Big)^{\dag}_{H_2V_2 H_1V_1}.
\end{equation}
A coincidence measurement in the same output of BS2 projects the optical modes onto the state $ \Big( | 0110 \rangle  + | 1001 \rangle  \Big)_{H_2V_2 H_1V_1}$, for example, corresponding to $\hat{M}_{+}$. This can be seen as follows:
\begin{equation}\label{eq6b}
\begin{split}
 \Big( | 0110 \rangle  &+ | 1001 \rangle  \Big)_{H_2V_2 H_1V_1} \\
 &=  (\hat{a}_V^{\dag} \hat{b}_H^{\dag} + \hat{a}_H^{\dag} \hat{b}_V^{\dag} ) \, | {\rm vac} \rangle   \\
 &= \frac{1}{2} \Big[ (\hat{c}_V^{\dag} - \hat{d}_V^{\dag}) (\hat{c}_H^{\dag} + \hat{d}_H^{\dag}) + (\hat{c}_H^{\dag} - \hat{d}_H^{\dag}) (\hat{c}_V^{\dag} + \hat{d}_V^{\dag})    \Big]\,  | {\rm vac} \rangle   \\
 &= (\hat{c}_H^{\dag} \hat{c}_V^{\dag}  -  \hat{d}_H^{\dag} \hat{d}_V^{\dag} ) \,   | {\rm vac} \rangle.
\end{split}
\end{equation}
Here $\hat{a}_H^{\dag}, \hat{a}_V^{\dag}, \hat{b}_H^{\dag}, \hat{b}_V^{\dag}$ are the creation operators for the optical modes $H_2, V_2, H_1, V_1$ at the input ports of BS2. Furthermore $\hat{c}_H^{\dag}, \hat{c}_V^{\dag}, \hat{d}_H^{\dag}, \hat{d}_V^{\dag}$ are the creation operators for the respective polarization output modes of BS2 (see Fig.~\ref{Fig2}) and $ | {\rm vac} \rangle $ denotes the vacuum state.

Similarly, the measurement $\hat{M}_-$ is realized by a coincidence in two detectors corresponding to orthogonal polarizations at different sides of BS2 according to
\begin{equation}\label{eq6c}
 \Big( | 0110 \rangle - | 1001 \rangle  \Big)_{H_2V_2 H_1V_1} \!\! = (\hat{c}_V^{\dag} \hat{d}_H^{\dag}  -  \hat{c}_H^{\dag} \hat{d}_V^{\dag} ) \,   | {\rm vac} \rangle.
\end{equation}  
Additionally, the other two types of Bell measurements $\Big( | 0101 \rangle \pm | 1010 \rangle  \Big)^{\dag}_{H_2V_2 H_1V_1}$ with coincidences in detectors corresponding to the same polarization require the detection of two-photon states $|2\rangle_{H/V}$. While this can be realized using photon-number-resolving detectors, we will disregard these cases here for simplicity and experimental feasibility.

A successful Bell-state measurement of the joint state $|\Phi \rangle$ of Eq.~\eqref{eq5} with $\hat{M}_+$ projects the two mechanical modes $A$ and $B$ into a state of the form
\begin{equation}\label{eq7}
|\psi'\rangle = \sin \theta \, |01 \rangle_{AB} + \cos \theta \, |10 \rangle_{AB}.
\end{equation}
Comparing this to the state in~\eqref{eq4}, we see that the only difference between the initial optical state $|\psi\rangle$ and the teleported mechanical state $|\psi'\rangle$ is that the probability amplitudes of the two eigenstates are exchanged. For the measurement $\hat{M}_-$, the two mechanical modes are projected onto the state
\begin{equation}\label{eq8}
|\psi'' \rangle = \sin \theta \, |01 \rangle_{AB} - \cos \theta \, |10 \rangle_{AB},
\end{equation}
which has a $\pi$-phase difference with respect to the state $|\psi' \rangle$.

In order to verify the successful teleportation, the mechanical state can be read out by sending a red-detuned pulse (i.e., utilizing the optomechanical state-swap interaction described in Sec.~\ref{tools}) and the teleported state can be analyzed in the Bell-state detection setup with a few additional optical components for fast polarization control. Alternatively, it could be routed elsewhere using an optical switch, enabling advanced quantum communication protocols. A phase difference in the final mechanical state from measurement with $\hat{M}_-$ can, in principle, be corrected in this readout step using a feed-forward operation by applying a phase shift in the optical interferometer.

\section{Experimental imperfections}

\subsection{Effect of residual thermal occupation}

For high-$Q$ mechanical oscillators the typical timescale at which our protocol can be realized is much shorter than the mechanical lifetime~\cite{Marinkovic2018,Riedinger2018,Wallucks2020}. Therefore, during a complete experimental run, the mechanics can be assumed to have negligible dissipation. In any real experiment, the mechanical modes are not perfectly initialized in their quantum ground state $|00 \rangle_{AB}$ and, in addition, are heated through optical absorption from the pulses interacting with the devices~\cite{Riedinger2016,Hong2017,Riedinger2018}. We model these effects by assuming the mechanical modes are {\it initially} in a thermal state with nonzero thermal occupation $\bar n_0 \ll 1$. Note that to simplify the model, we thus include the heating from the readout pulse into the initial residual thermal occupation as well. We furthermore assume that the two oscillators have equal thermal occupation $\bar n_0$, i.e., they are in the same thermal state,
\begin{equation}\label{eq9}
\rho_{th} = (1{-}s) \sum_{n=0}^{\infty} s^n \, | n \rangle \langle n | , 
\end{equation}
where $s\,{=}\,\frac{\bar n_0}{\bar n_0 +1} \,{<} \, 1$. For $\bar n_0 \,\,{<}\,\, 0.2$, $s \,\,{<}\,\, 0.17$, $s^2<0.03$, and $s^3<0.005$, high-excitation terms $|n \rangle$ ($n>2$) can be safely neglected and we approximate $\rho_{th} \simeq (1{-}s) \,  \Big(  | 0 \rangle \langle 0 | + s | 1 \rangle \langle 1 | +s^2 | 2 \rangle \langle 2 |  \Big)$. The density matrix of the two mechanical modes can then be written as $\rho_{th}^{AB} \, {\simeq}\,  \rho_{th}^{A}  \otimes  \rho_{th}^{B}$, which is a probabilistic mixture of nine pure states $| j k \rangle_{AB}$ ($j,k=0,1,2$). The eigenstate $| 00 \rangle_{AB}$ in $\rho_{th}^{AB}$ corresponds to the initial ground state assumed in Sec~\ref{ide}, which results in the mechanical state $|\psi' \rangle$ in~\eqref{eq7} ($|\psi'' \rangle$ in~\eqref{eq8}) after the Bell-state measurement $\hat{M}_\pm$. The remaining eight states eventually lead to unwanted additional terms in the final mechanical state and thus reduce the fidelity of the teleportation, which we will now analyze more closely.

\begin{figure}[t]\label{fig3}
\includegraphics[width=\linewidth]{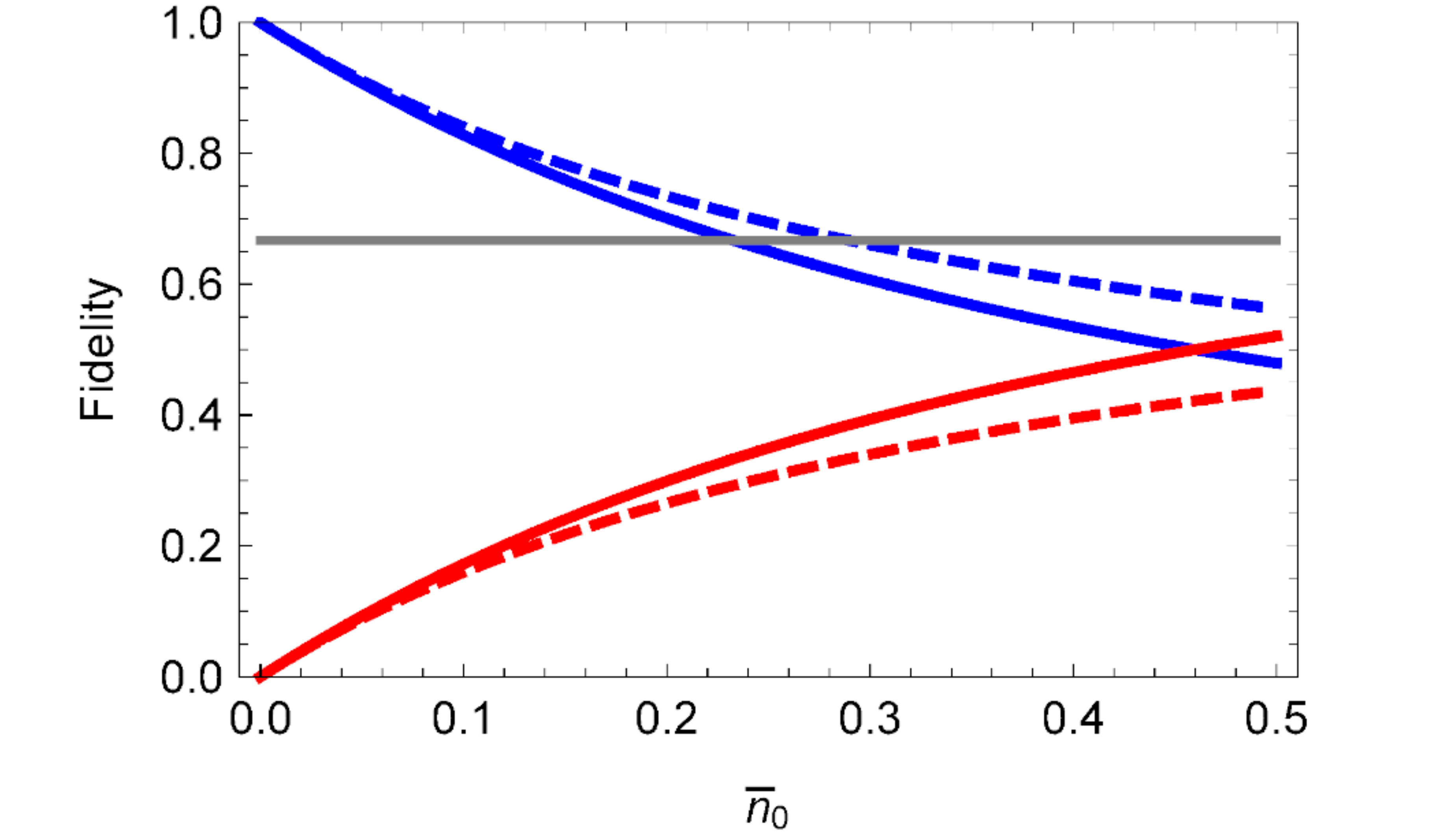}
\caption{The fidelity of the teleportation of the initial state $|\psi_{00} \rangle$ onto the final mechanical state $\rho_{final}^{AB}$ (top, blue lines), as well as the probability of unwanted additional terms (bottom, red lines) as a function of the initial mechanical thermal occupation $\bar n_0$. Solid (dashed) lines:\ approximation of neglecting $|n \rangle$, $n>2$ ($n>1$), terms in the state~\eqref{eq9}. Gray (horizontal, straight) line:\ minimum threshold of the state fidelity of 2/3 for quantum teleportation.}
\label{Fig3}
\end{figure}

For the mechanical modes initially in the thermal state $\rho_{th}^{AB}$, we obtain the following final (unnormalized) mechanical state after the Bell-state measurement $\hat{M}_+$:
\begin{equation}\label{eq12}
\begin{split}
\rho_{final}^{AB}  & \simeq  |\psi_{00} \rangle \langle \psi_{00} | + s |\psi_{01} \rangle \langle \psi_{01} | + s |\psi_{10} \rangle \langle \psi_{10} |  \\
& + s^2 |\psi_{11} \rangle \langle \psi_{11} |  +  s^2 |\psi_{02} \rangle \langle \psi_{02} | + s^2 |\psi_{20} \rangle \langle \psi_{20} | \\
& +  s^3 |\psi_{12} \rangle \langle \psi_{12} | + s^3 |\psi_{21} \rangle \langle \psi_{21} | + s^4 |\psi_{22} \rangle \langle \psi_{22} |, 
\end{split}
\end{equation}
where $|\psi_{jk} \rangle$ is the final mechanical state corresponding to the mechanical modes initially in the pure state $| j k \rangle_{AB}$ ($j,k=0,1,2$), and $|\psi_{00} \rangle \equiv |\psi' \rangle$. The remaining eight states $|\psi_{jk} \rangle$ are then given by
\begin{equation}
\begin{split}
|\psi_{01} \rangle &= \sin \theta \, |02 \rangle + \cos \theta \, |11 \rangle,   \\
|\psi_{10} \rangle &= \sin \theta \, |11 \rangle + \cos \theta \, |20 \rangle,   \\
|\psi_{11} \rangle &= \sin \theta \, |12 \rangle + \cos \theta \, |21 \rangle,   \\
|\psi_{02} \rangle &= \sin \theta \, |03 \rangle + \cos \theta \, |12 \rangle,   \\
|\psi_{20} \rangle &= \sin \theta \, |21 \rangle + \cos \theta \, |30 \rangle,   \\
|\psi_{12} \rangle &= \sin \theta \, |13 \rangle + \cos \theta \, |22 \rangle,   \\
|\psi_{21} \rangle &= \sin \theta \, |22 \rangle + \cos \theta \, |31 \rangle,   \\
|\psi_{22} \rangle &= \sin \theta \, |23 \rangle + \cos \theta \, |32 \rangle.   \\
\end{split}
\end{equation}
For the Bell-state measurement $\hat{M}_-$, we obtain $\rho_{final}^{AB}$ as in~\eqref{eq12}, but with the ``+" replaced with ``-" in each component $|\psi_{jk} \rangle$. Clearly, the above eight states are all unwanted contributions, as the total phonon number in the two mechanical modes is greater than 1 due to the residual thermal excitations in the initial state. These additional terms can be well suppressed if $\bar n_0 \ll 1$ (thus $s \ll 1$), as the total probability of the additional terms is given by
\begin{equation}\label{eq13}
P_{add} = \frac{2s+3s^2+2s^3+s^4}{1+2s+3s^2+2s^3+s^4}.
\end{equation}
We plot the probability of these terms for increasing $\bar n_0$ as red lines in Fig.~\ref{Fig3}. The probability of obtaining the ideal state $|\psi_{00} \rangle$, i.e., the fidelity of the state $\FF = \langle \psi_{00}  |\, \rho_{final}^{AB} \, |\psi_{00}  \rangle = 1-P_{add}$, thus decreases with increasing $\bar n_0$. A successful quantum teleportation requires the fidelity to be above $2/3$, which is given by the maximum achievable value without using entanglement as a resource~\cite{Massar1995}. For our optomechanical teleportation scheme, this corresponds to a threshold of $\bar n_0 \simeq 0.23$. For $\bar n_0 < 0.23$, the total probability of the higher-excitation terms $|n \rangle$ ($n>2$) in the thermal state~\eqref{eq9} is less than $0.7\%$, highlighting that our earlier approximation is quite good. In order to show optomechanical quantum teleportation, $\bar n_0 \simeq 0.23$ is therefore an upper bound in our present scheme.

\subsection{Weak coherent states input}

So far we have studied the special, idealized case where both optical input fields for the Bell-state measurement are single photons. This assumption, however, is unrealistic for devices in the optomechanical weak coupling regime which require blue-detuned pulses far above the single photon level to achieve practical scattering rates. Additionally, using a weak coherent state (WCS) as the optical input for the teleportation enables greatly reduced experimental overhead and can increase the rate of the experiment. While the single-photon case can be a good approximation for the case of low-excitation probabilities, we now include higher-excitation terms $|2 \rangle$ to the optomechanically entangled state, as well as the initial optical state $|\psi \rangle$. These terms are consequently affecting the final mechanical state, which results in corrections to the achievable teleportation fidelity. The WCS $|\alpha \rangle$ ($|\beta \rangle$) of the blue-detuned (resonant) pulse can be approximately expanded in the Fock-state basis as $|\RR \rangle \simeq | 0 \rangle + \RR | 1 \rangle + \frac{\RR^2}{\sqrt{2}}  | 2 \rangle $, which is a good approximation when $|\RR| \ll 1$, where $\RR=\alpha, \beta$.

We perform the analysis of the higher order terms considering several cases individually. We begin with the one where the resonant pulse $|\psi \rangle_{in}$ that is used to prepare the initial state $|\psi \rangle$ is a WCS but no optomechanical scattering occurs, i.e., $|\phi \rangle= |0000 \rangle_{AB H_2V_2}$. The single photon component of $|\psi \rangle_{in}$ will then not trigger any coincidences in the Bell-state detection, leading to unsuccessful trials for the teleportation. The two-photon component $| 2 \rangle$, however, results in an extra term in the final mechanical state. The joint state for the Bell-state measurement in this case is of the form
\begin{equation}\label{eq15}
|0000 \rangle_{AB H_2V_2} \otimes \Big( \sin^2 \!\theta \, |20 \rangle + \cos^2 \!\theta \, |02 \rangle +\frac{\sin 2\theta}{\sqrt{2}}  |11 \rangle  \Big)_{H_1V_1} .
\end{equation}
Such a state will trigger several different types of ``double-clicks", in particular,
\begin{equation}\label{eq15b}
\begin{split}
|0020 \rangle_{H_2V_2H_1V_1} \!\! &= \frac{1}{2\sqrt{2}} (\hat{c}_H^{\dag2} + \hat{d}_H^{\dag2} + 2 \hat{c}_H^{\dag} \hat{d}_H^{\dag} ) \,   | {\rm vac} \rangle,  \\
|0002 \rangle_{H_2V_2H_1V_1} \!\! &= \frac{1}{2\sqrt{2}} (\hat{c}_V^{\dag2} + \hat{d}_V^{\dag2} + 2 \hat{c}_V^{\dag} \hat{d}_V^{\dag} ) \,   | {\rm vac} \rangle,  \\
|0011 \rangle_{H_2V_2H_1V_1} \!\! &= \frac{1}{2} (\hat{c}_H^{\dag} \hat{c}_V^{\dag} + \hat{d}_H^{\dag} \hat{d}_V^{\dag} + \hat{c}_H^{\dag} \hat{d}_V^{\dag} + \hat{c}_V^{\dag} \hat{d}_H^{\dag} ) \,   | {\rm vac} \rangle.
 \end{split}
\end{equation}
Since successful Bell-state measurements of our protocol require coincidence detection of photons with orthogonal polarization, the double clicks in the same polarization modes induced by the states $ |20 \rangle_{H_1V_1}$ and $ |02 \rangle_{H_1V_1}$ can be identified and thus disregarded. The coincidences caused by the state $|11 \rangle_{H_1V_1}$, on the other hand, cannot be distinguished from those of $\hat{M}_{\pm}$ for the ideal case and result in errors in the final mechanical state. Fortunately, the mechanical modes are projected onto $|00 \rangle_{AB}$, which cannot be read out by the red-detuned pulse and thus has no actual impact on the measurement result. Note that in this section we assume zero thermal occupation of the mechanical modes, and focus on the effect of higher-order terms in the WCS of the blue-detuned and resonant pulses.

For the case where $|\psi \rangle_{in}$ is in a vacuum state $| 0 \rangle$, similarly, the single photon component in the blue-detuned pulse will not trigger double clicks and thus have no impact on the fidelity. The two-photon component $| 2 \rangle$ of this field, however, can again yield false coincidences in the Bell-state measurement and result in extra terms in the mechanical state. In this situation, the joint state at the input ports of BS2 is 
\begin{equation}\label{eq16}
\frac{1}{2} \Big(  |2020 \rangle + |0202 \rangle + \sqrt{2} |1111 \rangle  \Big)_{AB H_2V_2}  \otimes  |00 \rangle_{H_1V_1}.
\end{equation}
Similar to state~\eqref{eq15}, this state also triggers several different types of double clicks. The mechanical modes are then projected onto the states $|20 \rangle_{AB}$, $|02 \rangle_{AB}$, and $|11 \rangle_{AB}$ for optical modes being in $|20 \rangle_{H_2V_2}$, $ |02 \rangle_{H_2V_2}$, and $|11 \rangle_{H_2V_2}$, respectively. Since $|20 \rangle_{H_2V_2}$ and $ |02 \rangle_{H_2V_2}$ correspond to double clicks of the same polarization, such events can be identified and discarded, but $|11 \rangle_{H_2V_2}$ leads to double clicks that mix with those of Bell-state measurements $\hat{M}_{\pm}$ in the ideal case and result in an additional term $|11 \rangle_{AB}$ in the final mechanical state. The probability of this event with respect to the ideal case of the state $|\psi' \rangle$ ($|\psi'' \rangle$) is roughly  $|\frac{\alpha^2}{\sqrt{2}}|^2 : |\alpha \beta |^2 \,{=}\, \frac{1}{2} \frac{ |\alpha|^2}{ |\beta |^2} $. This implies that the effect of this additional term on the teleportation fidelity can be significantly reduced by using very weak blue-detuned pulses and introducing a hierarchy according to $|\alpha| \ll |\beta| \ll 1$.

For the other situations, where either of the two initial pulses is in $| 2 \rangle$, or one is in $| 1 \rangle$ and the other in $| 2 \rangle$, additional unwanted terms in the final mechanical state appear. However, their probabilities are much smaller compared to that of the ideal state $|\psi' \rangle$ ($|\psi'' \rangle$) since WCSs $|\alpha|, |\beta| \ll 1$ are used. The ratio of the probabilities is $\frac{|\beta|^2}{2} : \frac{|\alpha|^2}{2} : \frac{|\alpha|^2 |\beta|^2}{4} : 1$ for the two pulses in $|12 \rangle_{br}, |21 \rangle_{br}, |22 \rangle_{br}, |11 \rangle_{br}$, respectively, where the subscript $b$ ($r$) denotes the state of the blue-detuned (resonant) pulse.

\subsection{Photon loss and nonunity detection efficiency}

An experiment typically suffers from various optical losses including inefficient optical coupling to on-chip waveguides, passing through filter cavities, and photon detection with nonunity detection efficiency. These losses can be modeled by a beamsplitter (with transmittance $T$ and reflectance $R$, $T\,{+}\,R\,{=}\,1$), where the reflection is treated as optical losses~\cite{Leonhardt1997}. Given the different roles of optical losses, we classify them into two categories: detection losses and nondetection losses. 

We study the effect of nondetection or propagation losses on the final mechanical state in detail in the Appendix, considering the EPR state~\eqref{eq3} and the initial optical state~\eqref{eq4}. We show that for the Bell-state measurement $\hat{M}_+$, the unnormalized teleported state is of the form $\rho_{loss}^{AB} = T \, |\psi' \rangle \langle \psi' |$, where $0<T<1$ and $|\psi' \rangle$ is the final mechanical state~\eqref{eq7} in the ideal case without any losses. This implies that the optical losses will not affect the teleportation fidelity, but only increase the measurement time. This is also true for the other Bell-state measurement $\hat{M}_-$. We have further studied the case where the initial blue-detuned pulse is in a WCS, while the resonant pulse is a single photon. When the former is in a two-photon state $| 2\rangle$, the final mechanical state $\tilde{\rho}_{loss}^{AB}$ is a mixed state with eigenstates that are all unwanted with respect to $|\psi' \rangle$ (Appendix A). The probability of this state is much smaller than that of $\rho_{loss}^{AB}$ as they result from the states $| 2\rangle$ and $| 1\rangle$ in the expansion of WCS $|\alpha \rangle$, respectively, and the amplitude $|\alpha| \ll1$.

Similarly, the effect of detection losses in the Bell-state measurement setup can be analyzed. We know from~\eqref{eq6b} and~\eqref{eq6c} that without losses, $\hat{M}_{\pm}$ projects the optical modes onto Bell states $\Big( | 0110 \rangle  \pm | 1001 \rangle  \Big)_{H_2 V_2 H_1V_1}$, which correspond to different types of double clicks events. Detection losses can reduce them to either ``single-click" or ``no-click" events (see Appendix B), which then project the mechanical states onto mixed states. Since only double clicks are valid Bell-state measurements in our protocol, however, these can be disregarded such that there is no actual impact on the teleportation fidelity. Detection losses thus also only reduce the probability of obtaining the desired state, again implying longer measurement time.

\subsection{Additional imperfections}

Experiments using the above protocol are expected to suffer from additional imperfections. One of the hardest experimental challenges is to find identical optomechanical devices. Apart from a few exceptions~\cite{Pfeifer2016,Gil-Santos2017}, most systems to date have no in-situ tuning mechanism, such that devices with matching optical and mechanical frequencies have to be found after the fabrication is completed. Residual frequency offsets for devices which are well in the sideband-resolved regime mainly reduce the success probability for the teleportation by reducing the optomechanical scattering probabilities. Small additional offsets in the mechanical frequencies can be compensated for through the optical driving~\cite{Riedinger2018,Marinkovic2018}. Since the scattered fields are furthermore filtered before the Bell-state detection setup, such offsets do not affect the fidelity, but only the success probability. Dark counts on the detectors can also cause false detection events in the Bell-state measurement, reducing the fidelity of the teleported state. However, by using superconducting nanowire single-photon detectors with tens of Hertz of dark-count rates and accounting for pulse lengths of tens of nanoseconds, these typically play only a minor role.

\section{Conclusion}

We have presented a quantum teleportation protocol based on an optomechanical system, which teleports an unknown quantum state of an optical field onto a pair of massive mechanical oscillators. The protocol consists of two central steps: the generation of optomechanical EPR-type states, and Bell-state detection of the optical fields. We start from an idealized case to illustrate the essence of the protocol, where the optical pulses are single photons and the mechanical modes are in their quantum ground states. We then study various imperfections encountered in actual experiments. Specifically, we have analyzed the effects of residual thermal occupation, optical input of weak coherent states, photon losses, and nonunity detection efficiency. This allows us to find an optimal parameter regime in a realistic situation where quantum teleportation could be successfully demonstrated in a state-of-the-art optomechanics experiment. Realization of such a teleportation scheme may find applications in various quantum information subjects, such as quantum communication, quantum repeaters, and quantum computing, as well as for the fundamental study of macroscopic quantum phenomena.

\begin{acknowledgments}
We would like to thank Klemens Hammerer and Igor Marinkovi\'{c} for valuable discussions. This work is supported by the Foundation for Fundamental Research on Matter (FOM) Projectruimte grant (16PR1054), the European Research Council (ERC StG Strong-Q, 676842), and by the Netherlands Organization for Scientific Research (NWO/OCW), as part of the Frontiers of Nanoscience program, as well as through Vidi (680-47-541/994) and Vrij Programma (680-92-18-04) grants. R.B.\ and T.A.\ acknowledge funding from the Funda\c{c}\~{a}o de Amparo \`{a} Pesquisa do Estado de S\~{a}o Paulo (2019/01402-1, 2016/18308-0, 2018/15580-6) and from the Coordena\c{c}\~{a}o de Aperfei\c{c}oamento de Pessoal de N\'{\i}vel Superior (Finance Code 001). B.H.\ also acknowledges funding from the European Union under a Marie Sk\l{}odowska-Curie COFUND fellowship.
\end{acknowledgments}

After completion of this work, we became aware of a recent, related manuscript~\cite{Pautrel2020}.

\setcounter{figure}{0}
\renewcommand{\thefigure}{A\arabic{figure}}
\setcounter{equation}{0}
\renewcommand{\theequation}{A\arabic{equation}}

%\clearpage

%\section*{Supplementary Information}

\section*{Appendix: Effect of photon losses in optomechanical teleportation}

It is known that the process of linear optical loss can be modeled using a beamsplitter (BS), where its reflection is treated as the optical loss channel~\cite{Leonhardt1997}. This model allows us to conveniently obtain the state of an arbitrary quantum state after experiencing some loss. In this Appendix, we first study the effect of nondetection losses (such as loss in coupling optical pulses to mechanical modes, an optical mode passing through a filter, etc.) and derive the final mechanical state by including these losses. We then continue to study the effect of detection loss in a Bell-state measurement.

\begin{figure}\label{fig4}
\includegraphics[width=0.9\linewidth]{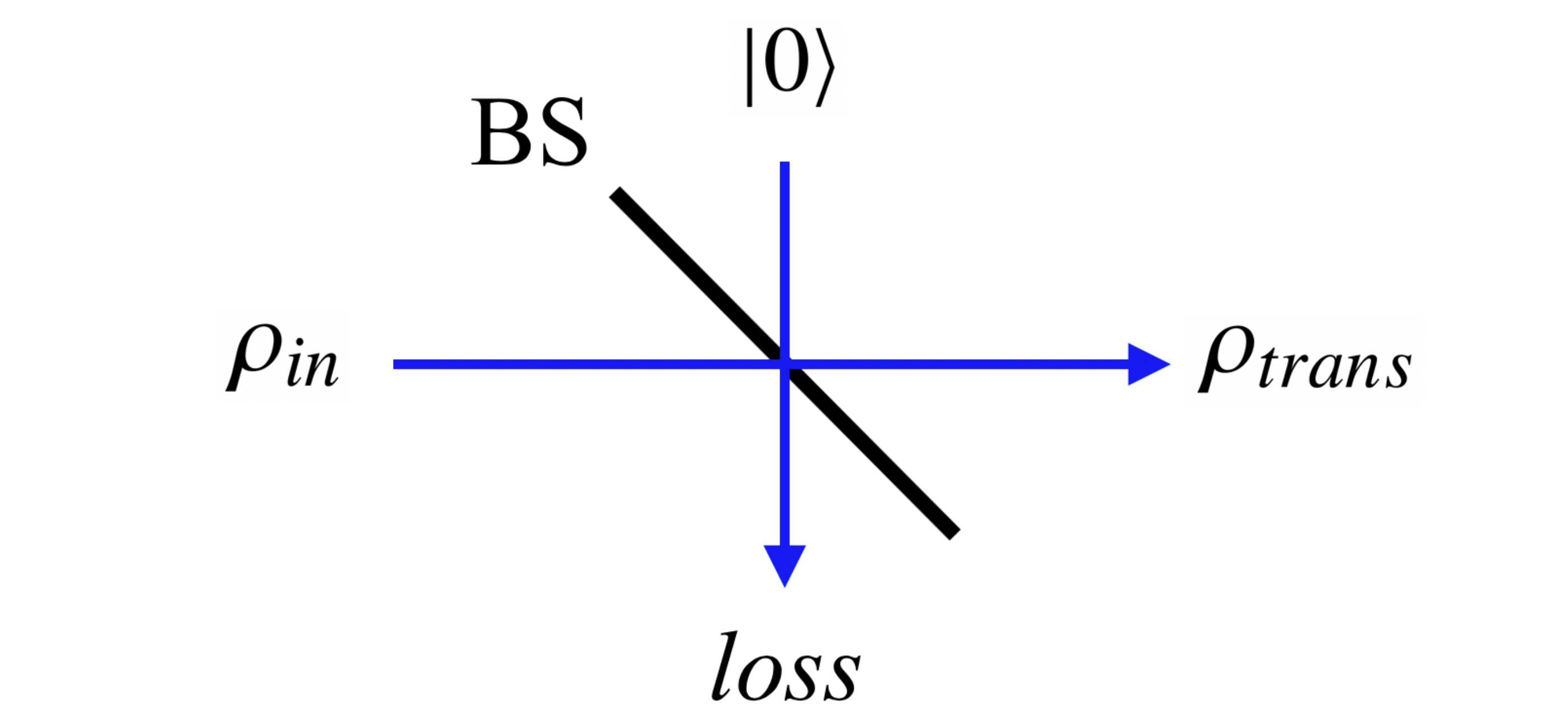}
\caption{Modeling of linear optical loss with a beamsplitter, where the reflection is treated as the loss. $\rho_{trans}$ denotes the state of an arbitrary quantum state $\rho_{in}$ after experiencing linear loss.}
\label{Fig4}
\end{figure}

For an arbitrary quantum state, denoted by a density matrix $\rho_{in}$, impinging on a BS and with vacuum entering through the other input port (cf.\ Fig.~\ref{Fig4}), the state of the two outputs can be expressed as~\cite{Ban1996}
\begin{equation}\label{a1}
\rho_{out} = \sum_{m=0}^{\infty}  \sum_{k=0}^{\infty} \Bigg[ \frac{1}{m!k!} \Bigg( \frac{R}{T} \Bigg)^{m+k}  \Bigg]^{\frac{1}{2}} \hat{a}^m T^{\frac{\hat{a}^{\dag} \hat{a}}{2}} \rho_{in}  T^{\frac{\hat{a}^{\dag} \hat{a}}{2}} (\hat{a}^{\dag})^k  \otimes | m \rangle \langle k |,
\end{equation}
where $\hat{a}$ and $\hat{a}^{\dag}$ are the annihilation and creation operators acting on the input state $\rho_{in}$, $T$ and $R$ are the transmittance and reflectance of the BS, respectively, and $T+R=1$ for a lossless BS. As any quantum state $\rho$ can be expanded in the Fock-state basis,
\begin{equation}
\rho= \sum_{n,s=0}^{\infty} C_{n,s} | n \rangle \langle s |,
\end{equation}
where $C_{n,s}$ are coefficients, we can replace $\rho_{in}$ with $| n \rangle \langle s |$ in~\eqref{a1} and trace over the lossy part (cf.\ Fig.~\ref{Fig4}). We then obtain the density matrix of the transmitted component~\cite{Li2010},
\begin{equation}\label{a2}
\rho_{n,s}= \sum_{m=0}^{{\rm min}(n,s)}  \sqrt{\frac{n!s!}{ (m!)^2 (n-m)! (s-m)! }  } R^m T^{\frac{n+s}{2}-m}  | n-m \rangle \langle s-m |,
\end{equation}
and therefore we can express an arbitrary state $\rho$ after experiencing linear optical loss as
\begin{equation}
\rho_{loss}= \sum_{n,s=0}^{\infty} C_{n,s} \, \rho_{n,s}.
\end{equation}
For a single-photon state $ |1 \rangle$, we have 
\begin{equation}
\rho_{loss}^{|1 \rangle}=  T | 1 \rangle \langle 1 | + R  | 0 \rangle \langle 0 |, 
\end{equation}
and for $| 0 \rangle \langle 1 |$ and $| 1 \rangle \langle 0 |$, we obtain
\begin{equation}
\rho_{loss}^{| 0 \rangle \langle 1 |} =  \sqrt{T} | 0 \rangle \langle 1 |,  \,\,\,\,\,\,\,\,\,   \rho_{loss}^{| 1 \rangle \langle 0 |} =  \sqrt{T} | 1 \rangle \langle 0 |.
\end{equation}

\subsection{Nondetection losses}
\label{AppendixA}

By placing a BS (which we call BS0) before the input port of BS2 in the main text (Fig.~\ref{Fig1}), we now include nondetection losses into the ideal case considered in Sec.~\ref{ide}. The optical components (i.e., two polarization modes $H_2$ and $V_2$) of the EPR state $|\phi \rangle$ in~\eqref{eq3} independently pass through BS0, and after tracing over the reflection, we hence obtain the following transmitted state, i.e., the state $|\phi \rangle$ after the inclusion of nondetection losses:
\begin{widetext}
\begin{equation}
\begin{split}
\rho_{loss}^{|\phi \rangle} &= \frac{1}{2} \Big[ |01\rangle_{AB} \langle 01| \otimes  |0\rangle_{H_2} \langle 0| \, \Big( T |1\rangle \langle 1| + R  |0\rangle \langle 0|  \Big)_{V_2}  +  |10\rangle_{AB} \langle 10 | \otimes \Big( T |1\rangle \langle 1| + R  |0\rangle \langle 0|  \Big)_{H_2}   |0\rangle_{V_2} \langle 0|    \\
& +  |01\rangle_{AB} \langle 10| \otimes \Big(\!\! \sqrt{T} |0\rangle_{H_2} \langle 1| \Big)  \, \Big(\!\! \sqrt{T} |1\rangle_{V_2} \langle 0| \Big)   +  |10\rangle_{AB} \langle 01| \otimes \Big(\!\! \sqrt{T} |1\rangle_{H_2} \langle 0| \Big)  \, \Big(\!\! \sqrt{T} |0\rangle_{V_2} \langle 1| \Big)   \Big]. 
\end{split}
\end{equation}
\end{widetext}
Combining this with the initial state $|\psi \rangle$ in~\eqref{eq4}, we therefore have the joint state at the input ports of BS2,
\begin{equation} 
\rho_{loss}^{|\Phi \rangle} = \Big[ \rho_{loss}^{|\phi \rangle} \Big]_{ABH_2V_2}  \otimes  \Big[ |\psi \rangle \langle \psi |  \Big]_{H_1V_1}  .
\end{equation}
After performing the Bell-state measurement $\hat{M}_+$ on the optical modes, consequently the mechanical modes are projected onto the unnormalized state,
\begin{widetext}\label{a3}
\begin{equation} 
\begin{split}
\rho_{loss}^{AB} &= \hat{M}_+ \, \rho_{loss}^{|\Phi \rangle} \hat{M}_+^{\dag} \\
&= T \Big(  \sin^2 \theta \, |01\rangle \langle 01|  +   \cos^2 \theta \, |10\rangle \langle 10|  +  \frac{ \sin 2\theta}{2} |01\rangle \langle 10| +  \frac{ \sin 2\theta}{2} |10\rangle \langle 01|    \Big)_{AB}   \\
&= T \, |\psi' \rangle \langle \psi' |, 
\end{split}
\end{equation}
\end{widetext}
where $|\psi' \rangle$ is the projected mechanical state~\eqref{eq7} in the ideal case without any losses. By comparing $\rho_{loss}^{AB}$ with $|\psi' \rangle$, we see that there is only a difference of a factor ``$T$", $0<T<1$, which means that the optical losses will not degrade the final mechanical state but only reduce its probability, implying longer measurement time.

The above analysis is for the ideal case of the optical pulses being single photon states. We now assume that the blue-detuned pulse is a WCS and derive the final mechanical state if this pulse is in a two-photon state $| 2\rangle$, while the resonant pulse is still in $| 1\rangle$. Following the same procedures, after extensive calculations, we obtain 
\begin{widetext}\label{a4}
\begin{equation} 
\begin{split}
\tilde{\rho}_{loss}^{AB} &= \frac{1}{4} \Bigg\{  \cos^2 \theta \, \Big[ \Big( T^2 + 2RT \Big) |20\rangle \langle 20| + 2RT |11\rangle \langle 11| \Big] + \sin^2 \theta \, \Big[ \Big(T^2 + 2RT \Big) |02\rangle \langle 02| + 2RT |11\rangle \langle 11| \Big]   \\
&+  \frac{ \sin 2\theta}{2} \Big[ T^2  |20\rangle \langle 02| + 2 RT \Big( |20\rangle \langle 11| + |11\rangle \langle 02| \Big)   \Big]  +  \frac{ \sin 2\theta}{2} \Big[ T^2  |02\rangle \langle 20| + 2 RT \Big( |02\rangle \langle 11| + |11\rangle \langle 20| \Big)   \Big]    \Bigg\}_{AB} .  \\
\end{split}
\end{equation}
\end{widetext}
This state contains eigenstates that are all unwanted with respect to the ideal state $|\psi' \rangle$. The probability of this state is much smaller than that of the state $\rho_{loss}^{AB}$ as they result from the states $| 2\rangle$ and $| 1\rangle$ in the coherent-state expansion, respectively. Therefore, a WCS $|\alpha| \ll1$ must be used in order to reduce the probability of those unwanted terms in the final mechanical state.

\begin{figure}\label{fig5}
\includegraphics[width=\linewidth]{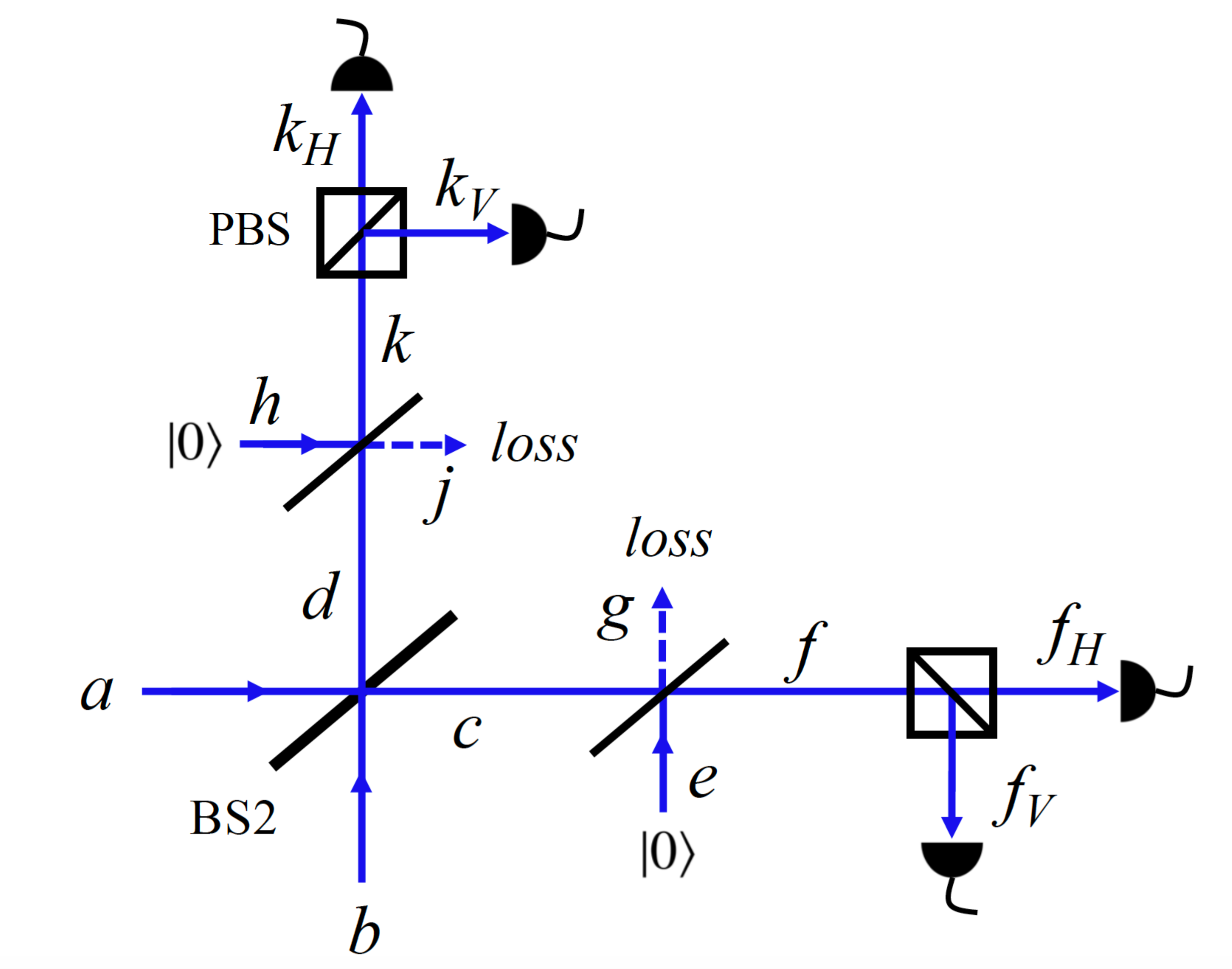}
\caption{Beamsplitter model of detection losses in the Bell-state measurement. The reflection of the hypo-BS in each arm of the outputs of BS2 models the detection loss. A single-mode operator is used to represent the two polarization modes, i.e., $\hat{O} \equiv \hat{O}_{H(V)}$.}
\label{Fig5}
\end{figure}

\subsection{Detection losses}
\label{AppendixB}

The nonunity detection efficiency in Bell-state measurements can be modeled by putting a BS in each arm of the two outputs of BS2, as depicted in Fig.~\ref{Fig5}. The reflections of the two hypothetical BSs (hypo-BSs) are used to model the detection losses, and the transmitted fields represent the optical fields after such losses and are measured by detectors with {\it unity} detection efficiency. We assume identical hypo-BSs given the symmetry of the Bell-state detection. The two output fields of BS2 then become the input fields of the hypo-BSs with vacuum entering the other input ports, and the transmitted fields of each hypo-BS are then measured by two detectors placed after a PBS, which measure orthogonal polarizations. Note that the transmitted fields contain two polarization modes and we use operator $\hat{O}_{H(V)}$ to denote the $H$ ($V$) polarization mode, where $\hat{O}=a,b,c,d,e,f,g,h,j,k$; see Fig.~\ref{Fig5}.  

In the ideal case without detection losses, we know from~\eqref{eq6b} and~\eqref{eq6c} that the Bell-state measurements $\hat{M}_{\pm}$ project the optical modes onto the state $ \Big( | 0110 \rangle  \pm | 1001 \rangle  \Big)_{H_2V_2 H_1V_1}$, which correspond to different types of double clicks in the four detectors shown in Fig.~\ref{Fig2}. We now study the effect of detection losses and see if the double clicks can still characterize the Bell-state measurements $\hat{M}_{\pm}$. 

The Bell-state measurement $\hat{M}_{+}$ projects the optical modes onto the state,
\begin{widetext}\label{d1}
\begin{equation}
\begin{split}
 \Big( | 0110 \rangle  &+ | 1001 \rangle  \Big)_{H_2V_2 H_1V_1} \\
 &= (\hat{c}_H^{\dag} \hat{c}_V^{\dag}  -  \hat{d}_H^{\dag} \hat{d}_V^{\dag} ) \,   | {\rm vac} \rangle  \\
 &= \Big[ \Big(\! \sqrt{T} \hat{f}_H^{\dag} -\! \sqrt{R} \hat{g}_H^{\dag} \Big) \Big(\! \sqrt{T} \hat{f}_V^{\dag} -\! \sqrt{R} \hat{g}_V^{\dag} \Big) -  \Big( \! \sqrt{T} \hat{k}_H^{\dag} -\! \sqrt{R} \hat{j}_H^{\dag} \Big) \Big( \! \sqrt{T} \hat{k}_V^{\dag} - \! \sqrt{R} \hat{j}_V^{\dag} \Big)    \Big]     \, | {\rm vac} \rangle   \\
 &= \Big( T |1100 \rangle  + R |0011 \rangle - \! \sqrt{TR} |0110 \rangle - \! \sqrt{TR} |1001 \rangle \Big)_{f_H f_V g_H g_V} \!\! - \Big( T |1100 \rangle  + R |0011 \rangle - \! \sqrt{TR} |0110 \rangle - \! \sqrt{TR} |1001 \rangle  \Big)_{k_H k_V j_H j_V}  \\
  &\equiv | \Psi_{out} \rangle. 
\end{split}
\end{equation}
\end{widetext}
Considering the detection losses, we trace over the output modes $g_{H(V)}, j_{H(V)}$ (which denote losses; see Fig.~\ref{Fig5}) and obtain 
\begin{widetext}\label{d2}
\begin{equation} 
\begin{split}
\rho_{fk} &= {\rm Tr}_{gj} \Big[ | \Psi_{out} \rangle \langle \Psi_{out} |  \Big]   \\
&=  \Big(  T^2 |11\rangle \langle 11|  +  R^2 |00\rangle \langle 00|  + TR  |01\rangle \langle 01|  + TR  |10\rangle \langle 10|   \Big)_{f_H f_V}   \\
&+ \Big(  T^2 |11\rangle \langle 11|  +  R^2 |00\rangle \langle 00|  + TR  |01\rangle \langle 01|  + TR  |10\rangle \langle 10|   \Big)_{k_H k_V} .  \\
\end{split}
\end{equation}
\end{widetext}
Clearly, the detection losses lead to the emergence of additional states, which correspond to either single-click or no-click events. Since we record only the event of double clicks, we actually measure the following state
\begin{equation}\label{d3}
\begin{split}
\rho'_{fk} =  T^2 \Big( |11\rangle_{f_H f_V} \langle 11|  + |11\rangle_{k_H k_V} \langle 11|  \Big),  
\end{split}
\end{equation}
which corresponds to double clicks in the two detectors at the same side of BS2. This is the same as in the case without losses (see~\eqref{eq6b}), but detection losses reduce the probability of successfully measuring the state as $T^2 <1$.

Similarly, for the other measurement $\hat{M}_{-}$, we actually measure 
\begin{equation}\label{d4}
\begin{split}
\rho''_{fk} =  T^2 \Big( |11\rangle_{f_V k_H} \langle 11|  + |11\rangle_{f_H k_V} \langle 11|  \Big) ,
\end{split}
\end{equation}
which corresponds to double clicks in the two detectors at two sides of BS2 that measure orthogonal polarizations. Comparing with~\eqref{eq6c} for perfect detectors, the probability of the successful double clicks event is reduced, again, resulting in longer measurement times.

As before, the above analysis is for the ideal case of the pulses being single-photon states. For the practical situation of pulses being WCSs, the components of high-excitation states will also lead to double clicks (errors) in the presence of detection losses. Nevertheless, when the coherent states are sufficiently weak, $|\alpha|, |\beta| \ll 1$, these errors are negligible.

%\bibliography{Mirror.bib}

%

\end{document}